\documentclass[twocolumn,secnumarabic,floatfix,amssymb,amsmath,nofootinbib,tightenlines,showpacs,superscriptaddress,aps,prb]{revtex4}
\usepackage{amsfonts}
\usepackage[colorlinks=true,linkcolor=blue,pagecolor=blue,filecolor=blue,menucolor=yellow,urlcolor=blue,citecolor=blue,anchorcolor=blue]{hyperref}%
\usepackage{graphicx}
\usepackage{dcolumn}
\usepackage{times}
\usepackage{sidecap}

\newcommand{\bq}{\begin{equation}}
\newcommand{\eq}{\end{equation}}

\newcommand{\sffn}{$S$/$F_1$/$F_2$/$N$}
\newcommand{\ff}{$F_1$/$F_2$}
\newcommand{\sn}{$S$/$N$}

\newcommand{\eg}{\textit{e.g. }}
\newcommand{\etal}{\emph{et al.}}

\begin{document}

\title{Meissner Effect Probing of
Odd-Frequency Triplet Pairing in Superconducting Spin Valves }

\author{Mohammad Alidoust }
\email{phymalidoust@gmail.com} \affiliation{Department of Physics,
Norwegian University of Science and Technology, N-7491 Trondheim,
Norway}

\author{Klaus Halterman}
\email{klaus.halterman@navy.mil} \affiliation{Michelson Lab, Physics
Division, Naval Air Warfare Center, China Lake, California 93555,
USA}

\author{Jacob Linder}
\email{jacob.linder@ntnu.no} \affiliation{Department of Physics,
Norwegian University of Science and Technology, N-7491 Trondheim,
Norway}
\date{\today}

\begin{abstract}
Superconducting correlations which are long-ranged in magnetic systems have attracted much attention due to their spin-polarization properties
and potential use in spintronic devices.
Whereas experiments have demonstrated the slow decay of such correlations, it has proven more difficult to obtain a smoking gun signature of their
odd-frequency character which is responsible, \eg for their gapless behavior. We here demonstrate that the magnetic susceptibility response of a
normal metal in contact with a superconducting spin-valve provides precisely this signature, namely in form of an anomalous \textit{positive}
Meissner effect which may be tuned back to a conventional \textit{negative} Meissner response simply by altering the magnetization configuration
of the spin-valve.
\end{abstract}


\maketitle

\section{introduction}

Superconducting spin valves with double ferromagnet layers \cite{oh}
(i.e., \ff) have very recently attracted interest due to the
appearance of long-range spin and triplet supercurrents when the two
$F$ layers have misaligned magnetizations \cite{trifu,rich,hik}. These
structures are of considerable interest, both experimentally for
being feasible to fabricate, and also because such long-range
Josephson currents were thought to occur only in systems with more
complicated magnetic inhomogeneity \cite{bergeret01, pajovic, alidoust, halasz, houzet,sperstad,crouzy,alidoust2}. Most
experimental efforts aimed at revealing the long-range triplet
correlations have measured the supercurrent \cite{norm,robinson, klose}, which
conveyed the long-range nature of the triplet correlations, but not
their intrinsic odd-frequency nature. We consider here a \sffn spin
valve configuration, where S represents the superconductor and $N$ a
normal nonmagnetic metal. For this type of spin valve, we show that
it is possible to utilize the Meissner response to unambigously
determine the existence of long-ranged triplet correlations {\it
and} their odd-frequency symmetry. This originates from  proximity
effects \cite{berg}, in which Cooper pairs in $S$ populate the adjacent normal
metal or ferromagnet regions \cite{klaus}. These induced superconducting
correlations respond to an applied magnetic field, ${H_a}$, by
setting up Meissner screening currents, which lead to observable
changes in the magnetic susceptibility, $\chi$.

The Meissner response was previously investigated
\cite{nar,hig,belz} as a way to elucidate proximity effects in
hybrid nonmagnetic \sn\;systems. More recently these types of
structures revealed peculiar proximity-induced reentrance effects in
$\chi$ with changing $H_a$ or temperature, $T$
\cite{mot1,mot2,mot3}. These effects were attributed to an
enhancement of the paramagnetic contribution to the susceptibility
\cite{bruder,lis,gog}, which in the diffusive regime, can depend on
the existence of electron-electron interactions in the $N$ layer
\cite{art}. Meissner effects in monodomain ferromagnets in contact with superconductors have also been considered \cite{bergeret, krawiec}. If there are spin-dependent interactions, including
those arising from sequences of $F$ layers with misaligned
magnetizations, a richer variety of proximity effects can emerge,
including the well known induced equal-spin odd-frequency triplet
correlations with long-range penetration into a ferromagnet. Spin
dependent scattering at an \sn\;interface can also generate
odd-frequency correlations that modify the Meissner response,
turning it paramagnetic \cite{yoko}, resulting in  $\chi$
oscillations as $T$ is varied. A vanishing Meissner response may be
indicative of the onset of the familiar damped oscillations that a
Cooper pair undergoes when entering a ferromagnet \cite{miro}. It is
therefore of importance to understand not only how the odd triplet
pairing extend throughout the system, as described by the anomalous
Green function, but also to know their symmetry. There have been
considerable efforts to find ways to experimentally measure
the symmetries or long range nature of triplets, including surface
impedance measurements \cite{asano}, and local signatures in the
density of states (DOS) \cite{bergdos,halterman_08,cot,kaw,jac,yokoyama_07,annunziata}.

Recent experiments \cite{lek} involving
double
magnet superconducting spin valves 
have shown
that by varying
the relative
in-plane
magnetization angle, $\theta$,
the critical temperature, $T_c$,  can be
lowest when the magnetizations are
nearly orthogonal-reflecting the increased presence of equal-spin
triplet pairs \cite{lek2}, in agreement also with theoretical works
\cite{gol1,gol2,kh}. The long-range triplet pair correlations always
vanish when the magnetizations are collinear but can, due to phase
shifts at the \ff\;interface, oscillate and vanish at an intermediate
$\theta$ \cite{gol2}. By considering a \sffn spin valve,
manipulating $\theta$ controls the long-range odd-frequency triplets
that propagate in the $N$ region. Thus, when  the long-range triplet
correlations are absent or very weak, we show that the magnetic
susceptibility is negative, corresponding to the conventional
Meissner response. However, when there is a strong misalignment of
the mutual magnetizations, and the long-ranged triplets are
enhanced, we find that the magnetic susceptibility  is positive,
corresponding to an anomalous Meissner response. Therefore by
measuring $\chi$, it is possible to determine from its magnitude the
presence of the long-ranged correlations, while its sign indicates
their odd-frequency character. A main advantage of this prediction compared to previous works is that
the odd-frequency symmetry is revealed by an overall sign change which is easily experimentally observable, in contrast to more subtle
signatures such as scaling behavior and a combination of multiple spectroscopic fingerprints \cite{cot,kaw}.

\section{Theory and formalism} 
\begin{figure}[t!]
\includegraphics[width=6.50cm,height=3.0cm]{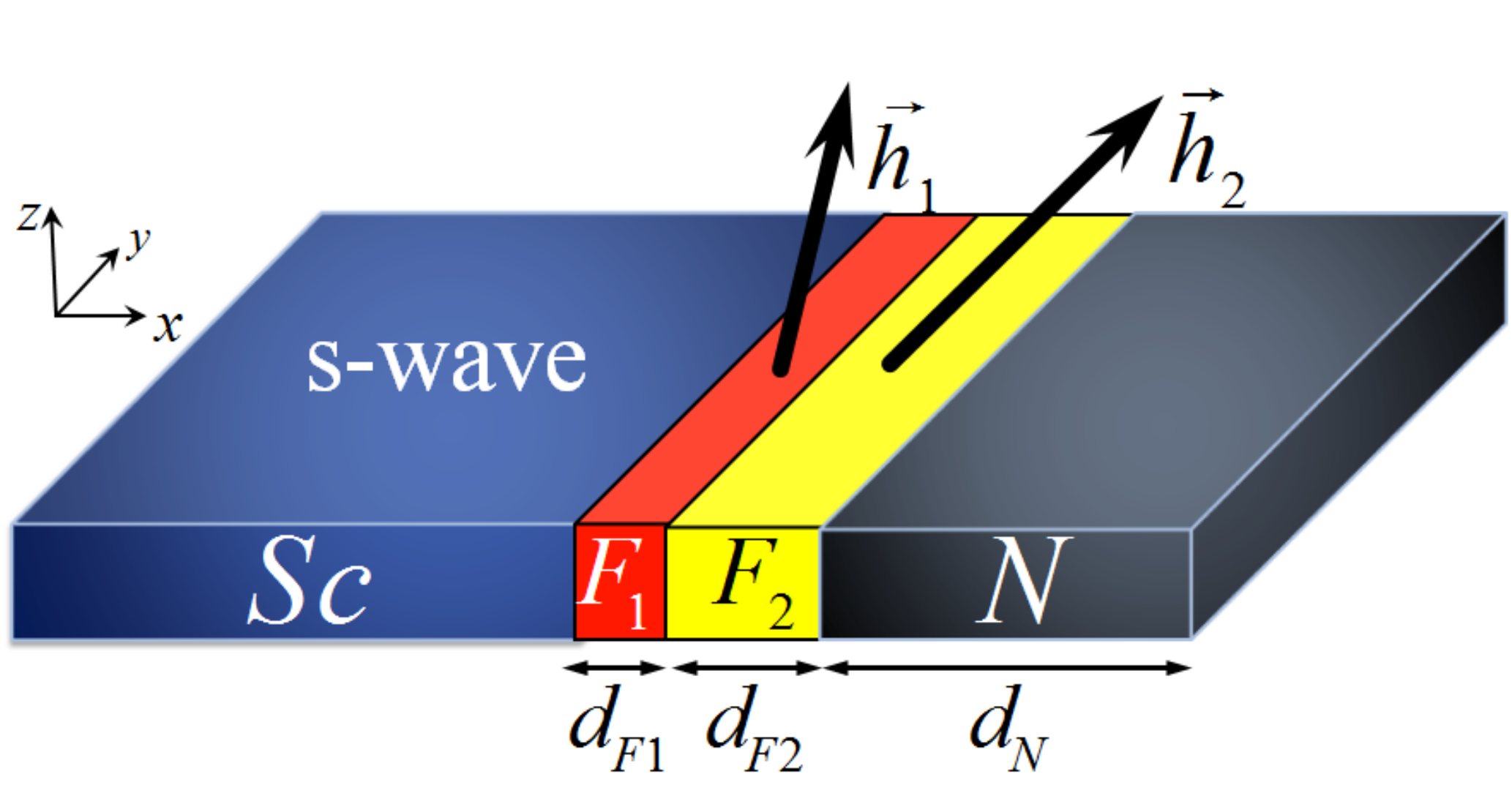}
\caption{\label{fig:model}(Color online) Schematic of the proposed
metallic \sffn junction. The $x$ axis is perpendicular to the
interfaces separating the different layers. Two ferromagnetic layers
($F_1$, $F_2$) with thicknesses $d_{F1}$ and $d_{F2}$ are sandwiched
between a superconductor ($S$) and normal metal ($N$) with thickness
$d_N$. The exchange field in the ferromagnetic layers, $\vec{h}_1$
and $\vec{h}_2$, may be misaligned as shown in the figure. An
external magnetic field $\vec{H}_a$ (not shown) is applied in the
$z$ direction, resulting in a spatially dependent Meissner
supercurrent $\vec{J}(x)$ flowing in the $y$ direction.}
\end{figure}

To model a realistic experimental
system, we will consider the diffusive limit for a metallic \sffn
junction as shown in Fig. \ref{fig:model}. The superconducting
proximity effect in such a setup may be described by using a Green's
function approach, where the superconducting correlations are
quantified via the so-called anomalous Green's function $\hat{f}$.
In the presence of a magnetic exchange field, it is necessary to
consider carefully the spin-structure of the Green's function which
in general will consist of a spin-singlet and spin-triplet part. In
order to compute the Meissner response of the normal metal, we first
need to obtain the anomalous Green's function in the normal part of
the junction. This is done by solving the quasiclassical Usadel
equation with proper boundary conditions at the interface region of
the superconductor ($x=0$) and vacuum $(x=d_{F1}+d_{F2}+d_N)$. In
typical experiments, the interface transparency is rather low
(tunneling limit) such that the linearization of the Usadel equation
is a very good approximation, corresponding to an anomalous Green's
function which satisfies $|\hat{f}|\ll 1$. We decompose
the anomalous Green's function into
spin-zero
($\mathbb{S}$) and spin-one triplet ($\vec{\mathbb{T}}$)
components as
\begin{align}
\hat{f}(\varepsilon)=
i(\mathbb{S}(\varepsilon)+\vec{\mathbb{T}}(\varepsilon).\vec{\tau})\tau_y.
\end{align}
Above, $\vec{\tau}$ is a vector composed of
Pauli matrices. The linearized Usadel equations then have the general form (with $\sigma=\pm1$):
\begin{align}\label{Linearized Usadel Eq.}
-\sigma\partial_x^{2} \mathbb{T}_{x-} &+i\partial_x^{2} \mathbb{T}_{y-} + 2i[- \varepsilon (-\sigma \mathbb{T}_{x-}+i \mathbb{T}_{y-}) \notag\\
&-\sigma \mathbb{S}_{-} (h_{x} -\sigma i h_{y}) ]=0,\notag\\
-\sigma\partial_x^2\mathbb{S}_{-} &+\partial_x^2\mathbb{T}_{z-} + 2i[-\sigma h_{x}\mathbb{T}_{x-}  -\sigma  h_{y}\mathbb{T}_{y-}  \notag\\
&-  (-\sigma\mathbb{S}_{-}+\mathbb{T}_{z-}) (\varepsilon+ \sigma h_{z}) ]=0.
\end{align}
Here, $D$ is the diffusion constant, $\vec{h}=(h_x,h_y,h_z)$ is the magnetic exchange field which is non-zero inside the ferromagnetic layers and absent in the normal metal layer.
The above equations are to be supplemented with the Kupriyanov-Lukichev boundary
conditions at $S$/$F_1$ interface and are valid in the tunneling
limit \cite{kup}.
The transparency of the interface to quasiparticle tunneling  is determined by the
parameter $\zeta$ which depends on the
the ratio between the resistance of the interface and the resistance
in the diffusive normal region.
We obtain the following boundary conditions at the $S$/$F_1$
interface:
\begin{eqnarray}\label{bc_SF1}
&&(\zeta\partial_x  - c^{\ast}(\varepsilon))(-\sigma \mathbb{T}_{x-}+i \mathbb{T}_{y-})=0,\\
&&(\zeta\partial_x  - c^{\ast}(\varepsilon))(-\sigma\mathbb{S}_{-}+\mathbb{T}_{z-}) +\sigma s^{\ast}(\varepsilon) =0.
\end{eqnarray}
Here $\mathbb{T}_{x,y,z\mp}\equiv\mathbb{T}_{x,y,z}(\mp\varepsilon)$,
whereas $s(\varepsilon)$, $c(\varepsilon)$ are the off-diagonal and normal components of the bulk Green's function
for the superconducting region. The general expression for the supercurrent density in
the system reads ${\vec{J}}\text{(}\vec{r}\text{)}=\vec{J}_{0}\int
d\varepsilon\text{Tr}\{\rho_{3}(\hat{G}[\hat{\partial},\hat{G}])^{K}\}$,
in which $K$ represents the Keldysh component of the matrix. Here $J_{0}$ is a normalization constant. When
the normal part of the system is subject to an external magnetic field $H_a$
oriented in the $z$ direction (see Fig. \ref{fig:model}) the
supercurrent flowing parallel to the interfaces (along $y$) can be expressed in
terms of the short-range spin-0 correlations and the long-range spin-1 triplet correlations
as:
\begin{align}\label{eq:current}
 J(x) &=-J_08ieA(x) \sum_\sigma\int_0^{\infty} d\varepsilon [
\sigma\mathbb{S}(\sigma\varepsilon,x)\mathbb{S}^{\ast}(-\sigma\varepsilon,x) \notag\\
&- \sum_{j=\{x,y,z\}} \sigma\mathbb{T}_j(\sigma\varepsilon,x)\mathbb{T}_j^{\ast}(-\sigma\varepsilon,x)]
\tanh(\varepsilon\beta/2).
\end{align}

Since the magnetic field $\vec{B}$
and the vector potential $\vec{A}$ are related by $\vec{\nabla}\times \vec{A}=\vec{B}$, Maxwell's equation
must be taken into account together with the above equation for the
supercurrent. If we use the Coulomb gauge,
$\vec{\nabla}\cdot\vec{A}=0$, Maxwell's equation reduces to
\begin{align}
d^2A(x)/dx^2=-4\pi J(x)/c.
\end{align}
Similar to previous works considering the proximity-induced Meissner effect \cite{yoko}, we assume that the
external magnetic field is expelled entirely from the superconductor and thus
$A(0)=0$. The normal metal is taken to be sufficiently wide so that the
external magnetic field fully penetrates the right-most end
of it. The system susceptibility is expressed by:
\begin{eqnarray}
\chi=\frac{1}{H_a d_N}\int M(x)dx=\frac{1}{4\pi H_a d_N}\int (B-H_a)dx,
\end{eqnarray}
where $\vec{B}=\vec{H}_a+4\pi \vec{M}$, and $\vec{M}$ is the magnetization.

\begin{figure*}
 \centering
\includegraphics[width=16.80cm,height=6.0cm]%
{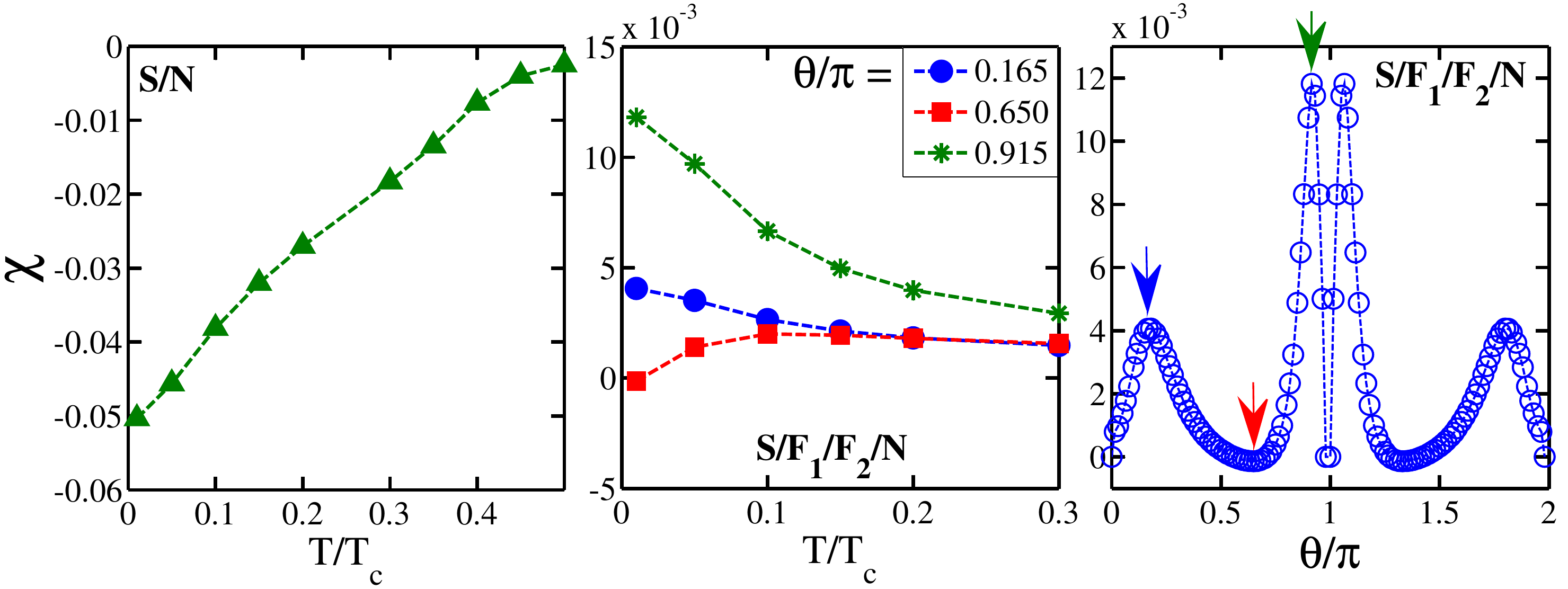} \caption{(Color online) Meissner effect in \sn
(left panel) and \sffn (middle and right panels) structures as a
function of temperature $T$ and magnetic misalignment angle
$\theta$. In the \sn structure the singlet superconducting
correlations give rise to a conventional Meissner effect ($\chi<0$)
while in the \sffn structure an anomalous Meissner effect ($\chi>0$)
appears which may be controlled via the relative magnetization
directions in the double F layers.} \label{fig:chi_T_thet}
\end{figure*}

\section{Results and Discussion} We have solved numerically the
Usadel equations with their boundary conditions, computed the
supercurrent density, and used the latter to solve Maxwell's
equation in order to obtain the magnetic susceptibility. This
corresponds to a linear-response theory since $J(x) \propto A(x)$,
which nevertheless compares very well to experiments \cite{belz}. To
enhance the level of induced triplet correlations in our system, the
thickness of one magnet should be much larger than the other one
\cite{trif}: the thin F layer generates the triplet components
whereas the thick F layer filters out the short-range components. We
thus set $d_{F1}=0.15\xi_S$, $d_{F2}=1.95\xi_S$, and $d_N=2.5\xi_S$
where $\xi_S$ is the superconducting coherence length, which is
typically of order tens of nm in diffusive metals. For simplicity,
the exchange field is taken to reside solely in the $y-z$ plane, so
that the angle $\theta$ describes the in-plane misalignment between
the two magnets. To ensure that the linearized Usadel equations are
valid, we consider the tunneling limit and set $\zeta=6$ whereas the
exchange field is taken to be that of a weak ferromagnetic alloy,
$|\vec{h}_1|=|\vec{h}_2|=10\Delta_0$. All lengths in the system are
normalized by the superconducting coherence length $\xi_S$ and
correspondingly, all energies by the superconducting gap at zero
temperature $\Delta_0$. Throughout the calculations we set that
$\hbar=k_B=1$, and unless otherwise noted,  $T=0.05T_c$.

To make contact with previous works considering non-magnetic \sn bilayers, we plot in the left panel of Fig. \ref{fig:chi_T_thet} the proximity-induced susceptibility $\chi$ integrated over the entire N region with $|\vec{h}_1|=|\vec{h}_2|=0$. As seen, the temperature-dependence of $\chi$ displays a diamagnetic behavior as expected for the conventional Meissner effect and vanishes as $T\to T_c$ \cite{belz}. We now investigate what happens when the spin-valve is present and denote the misalignment angle between the F layers as $\theta$. The middle panel of Fig. \ref{fig:chi_T_thet} gives the temperature-dependence of $\chi$ for an \sffn junction when considering different values of $\theta$. It is clear that the susceptibility changes sign compared to the non-magnetic case, indicating a paramagnetic Meissner effect. As we shall demonstrate below, this anomalous magnetic response is entirely due to the odd-frequency nature of the triplet correlations.

We have also computed how $\chi$ changes as a function of the misalignment angle $\theta$ as it makes a full cycle starting from a parallel magnetization configuration (P, $\theta=0)$ to antiparallel (AP, $\theta=\pi)$ and then back to parallel $(\theta=2\pi)$. This is shown in the right panel of Fig. \ref{fig:chi_T_thet} which displays several noteworthy features:
First, it is seen that $\chi$ practically vanishes at $\theta=0$ and $\theta=\pi$.
This is consistent with the fact that in these cases a single quantization axis can be defined for the whole system and hence
there
are no long-ranged odd-frequency triplets for these configurations,
reflecting the absence of proximity-effects. Secondly, two
peaks develop as soon as one moves slightly away from the P and AP configuration whereas $\chi$ has a minimum
and nearly vanishes for an intermediate value of $\theta$ near $\pi/2$. This might seem odd at first glance. Indeed,
one might expect the triplet correlations to largest in magnitude when the misalignment angle deviates the most from
the P and AP configuration, namely at $\pi/2$.

Before resolving this issue, we demonstrate that the anomalous
Meissner effect shown in Fig.~\ref{fig:chi_T_thet} is due to the
odd-frequency correlations induced in the normal metal. By directly
computing the spin-0 singlet and $S_z=0$ triplet contribution
$\mathbb{T}_z$ to the current in Eq. (\ref{eq:current}), we find
that these are several orders of magnitude smaller (and nearly zero)
compared to the contribution from $\mathbb{T}_y$. The latter is
exactly the long-ranged triplet component which must have an
odd-frequency symmetry since our system is diffusive (only $s$-wave
symmetry survives). Now, one may show analytically \cite{yoko} that
in the scenario where purely odd-frequency correlations are present,
the Meissner response will be paramagnetic such that $\chi>0$.
Therefore, the behavior of $\chi$ in our system provides a clear
signature not only for the long-ranged proximity effect, but
\textit{simultaneously} for its odd-frequency character. Another
advantage with the proposed setup is that one may explicitly tune
the Meissner response by changing the misalignment angle $\theta$,
since the latter turns on and off the long-ranged odd-frequency
correlations. We note that the magnitude of the proximity-induced
susceptibility $\chi$ $(\sim 0.01)$ is due to the assumption of a
weak proximity effect, and is well within experimental reach
\cite{visa,belz}. 

\begin{figure}[b!]
 \centering
\includegraphics[width=8.0cm,height=4.6cm]%
{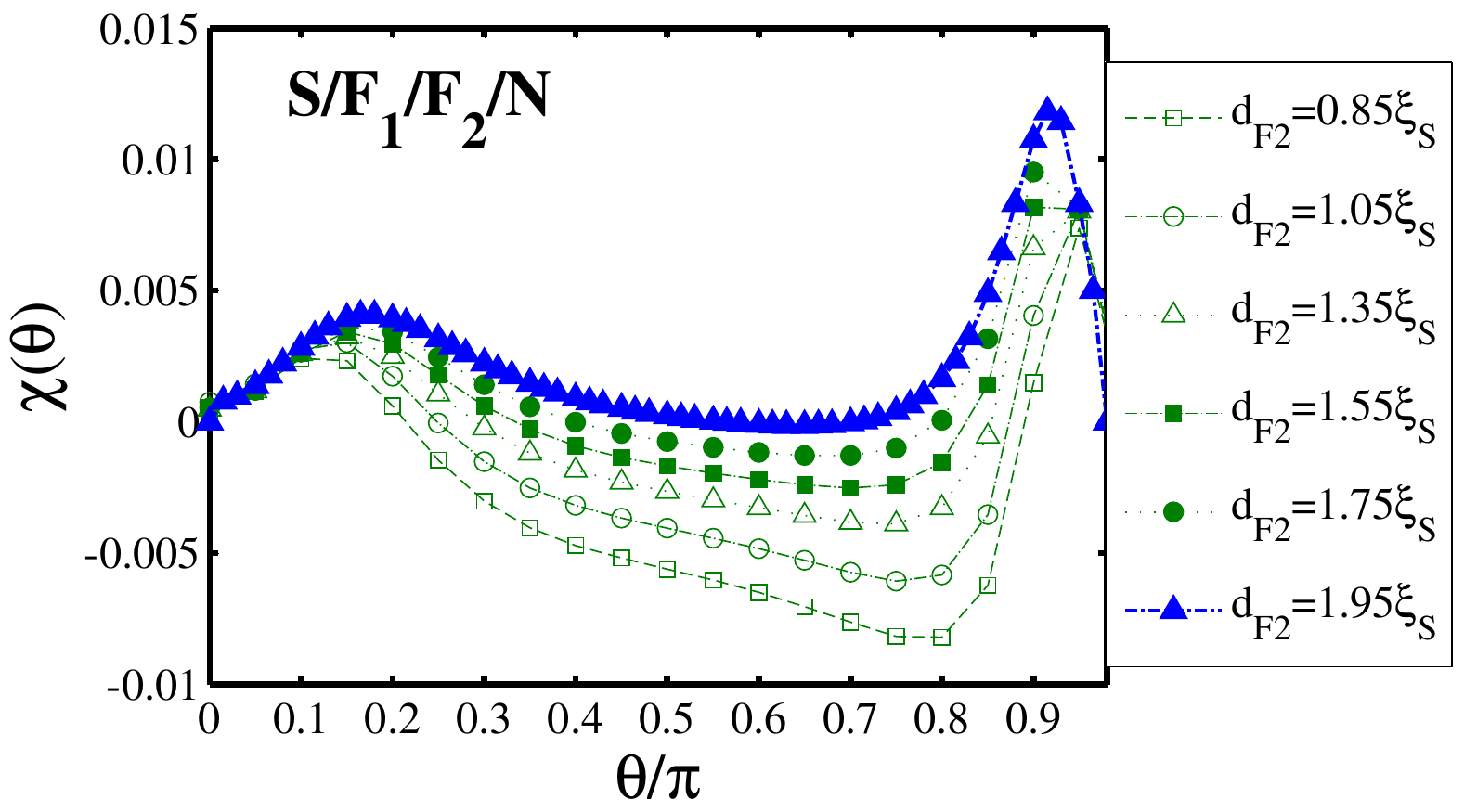} \caption{(Color online) Meissner effect ($\chi$) in
\sffn
 connections against the non-collinearity angle $\theta$ for various values of $F2$ layer thicknesses, $d_{F2}/\xi_S=0.85, 1.05, 1.35, 1.55, 1.75, 1.95$.
The thickness of $F1$ and normal metal layers are fixed at
$d_{F1}=0.15\xi_S$, and $d_{N}=2.5\xi_S$, respectively. The
anomalous Meissner effect ($\chi>0$) also depends on the magnetic
layer thicknesses and is strongest where $d_{F2}\gg d_{F1}$, for
instance.} \label{fig:chi_vs_df}
\end{figure}

Figure \ref{fig:chi_vs_df} illustrates the
crossover between a fully and partially anomalous Meissner effect,
$\chi>0$, as a function of magnetization misalignment $\theta$,
for several choices of the thickness of $F2$. The thickness of $F1$ is
kept fixed at $d_{F1}=0.15\xi_S$ whereas the thickness of second
ferromagnetic layer varies from $d_{F2}=0.85\xi_S$ to $d_{F2}=1.95$
where $d_{F1}\ll d_{F2}$. As seen, the anomalous Meissner
effect is most pronounced ($\chi>0$ for all $\theta$) when $d_{F1}\ll d_{F2}$. When $d_{F2}$ is reduced, the contribution to the Meissner effect from the short-ranged correlations, including the conventional even-frequency ones that yield a standard Meissner effect, is no longer negligible and thus $\chi$ becomes negative for an increased regime of misalignment angles $\theta$.

\begin{figure}[t!]
\includegraphics[width=8.20cm,height=4.2cm]{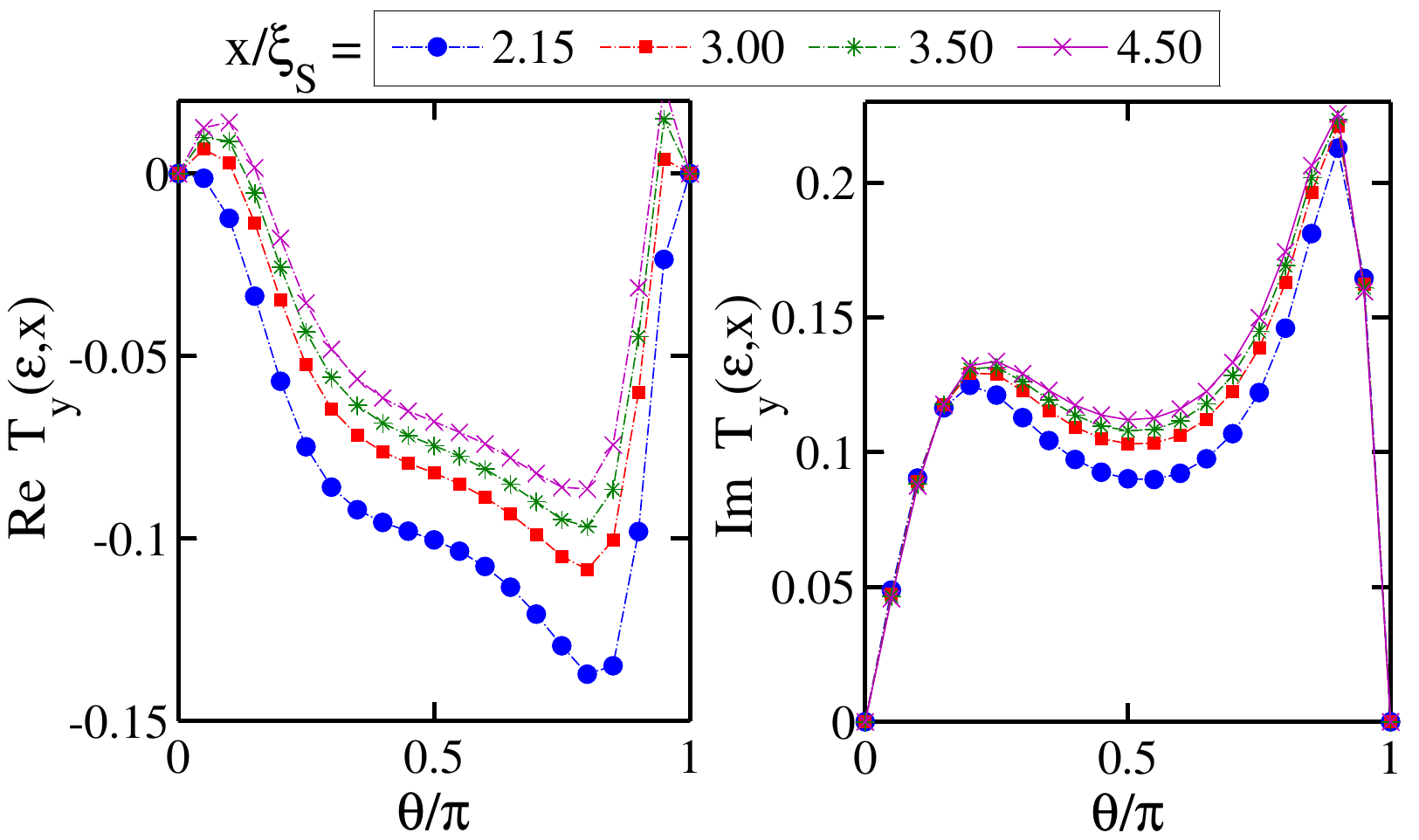}
\caption{\label{fig:gf}(Color online) Real and imaginary parts of the $y$-component of
the decomposed spin Green's function ($\mathbb{T}_{y}(\varepsilon,x)$)
versus magnetization misalignment angle $\theta$ at several locations
inside the $N$ region of the \sffn configuration considered here. Since
energies close to the Fermi level comprise the main contribution to the proximity effect, we have
set $\varepsilon$=0.1$\Delta_0$ as a representative choice. }
\end{figure}

Having established this, we now turn to the peculiar behavior near
$\pi/2$ for the susceptibility. The key thing to note here is that
contrary to what one might expect initially, the generation of
triplet correlations is \textit{not} necessarily the strongest for a
maximum misalignment angle near $\pi/2$. In fact, the
proximity-induced equal-spin triplets can actually vanish all
together for $\theta$ near $\pi/2$. This was recently shown in Ref.
\onlinecite{gol2} and explained in terms of phase-slips of the
singlet and triplet correlation functions due to the finite $S/F_1$
interface transparency, causing them to vanish at a critical
misalignment angle $\theta_c$ which in the tunneling limit and for
$h\gg\Delta_0$ was close to $\pi/2$. To investigate if a similar
situation is present in our setup, we plot in Fig. \ref{fig:gf} the
imaginary and real part of the long-ranged spin-triplet Green's
function $\mathbb{T}_{y}(\varepsilon,x)$. We consider  quasiparticle
energies close to the Fermi level, $\varepsilon/\Delta_0=0.1$. This
is a representative choice since the main contribution to the
current in Eq. (\ref{eq:current}) comes precisely from low-energy
excitations. Fig. \ref{fig:gf} shows how the anomalous Green's
function depends on the misalignment angle for several positions in
the $N$ metal: $x/\xi_S=\{2.15, 3.00, 3.50, 4.50\}$. As seen, the
anomalous Green's function displays qualitatively a similar
dependence on $\theta$ as the susceptibility, with peaks occurring
close to 0 and $\pi$ and a local minimum near $\pi/2$. This is then
consistent with the phase-slip phenomenon in Ref. \onlinecite{gol2}
and explains the angular dependence of $\chi$. The misalignment
angle $\theta$ then controls the long-ranged proximity effect in the
N part, which in turn alters the magnitude of the Meissner response
- the \textit{sign} of which is nevertheless always positive due to
the odd-frequency character of the correlations.

What is the physical origin of the paramagnetic Meissner effect
encountered here which serves as a signature for long-ranged
odd-frequency pairing? To explain this, it is useful to recall that
a non-linear Meissner effect also is known to occur in $d$-wave
superconductors \cite{prus}. In that case, the
temperature-dependence of the magnetic penetration depth has been
experimentally observed to be non-monotonic with a minimum at an
intermediate temperature between $T=0$ and $T=T_c$. This is in
contrast to what one sees for a conventional Meissner effect, which
simply reduces the penetration depth as the temperature decreases,
consistent with the left panel of our Fig. \ref{fig:chi_T_thet}. The
explanation behind this phenomenon for $d$-wave superconductors is
the existence of low-energy Andreev bound states which have a
paramagnetic contribution to the Meissner effect, thus competing
with the shielding supercurrent and producing the non-linear
behavior \cite{fog}. The key observation is now that Andreev-bound
states and odd-frequency pairing are intimately related, as shown in
Ref. \onlinecite{tanaka1}: in fact, the appearance of zero-energy
bound states in unconventional superconductors may be reinterpreted
as a manifestation of odd-frequency pairing correlations, such that
the latter should also contribute paramagnetically to the Meissner
effect. We nevertheless emphasize that in our case, the anomalous
Meissner effect is produced from garden-variety $s$-wave
superconductivity combined with a spin-valve structure, i.e. without
the need of any unconventional superconducting materials.

One experimental challenge regarding measurements of the predicted paramagnetic Meissner effect occurring in superconducting spin-valves comes from the influence of stray fields from the ferromagnetic regions. However, we note that the positive $\chi$ appears with a similar dependence on the misalignment angle $\theta$ even if one considers the susceptibility as obtained from integration over the entire non-superconducting region, i.e. $F_1/F_2/N$. The response of the bulk superconducting region is diamagnetic and could be expected to dominate the total susceptibility if one were to measure the susceptibility response of the entire heterostructure. In order to prove the combined odd-frequency and long-ranged nature of the superconducting proximity effect, it would therefore be necessary to conduct a local measurement of the susceptibility. As mentioned earlier, a measurement of the susceptibility response of the normal part would provide the most clear signature of the anomalous Meissner effect, although it would also be seen even if one were to include the ferromagnetic parts in the measurement. Major experimental advances in achieving such local probing of spin susceptibility has recently been reported, demonstrating detection capability at the single-spin level on nano- and atomic scales by using so-called nitrogen-vacancy magnetometers \cite{grinolds}. The observation of a paramagnetic Meissner effect in the spin-valve setup of Fig. \ref{fig:model} would provide clear evidence of not only the long-ranged nature of the triplet correlations, but importantly their odd-frequency characteristic, the latter not being observable in supercurrent measurements done so far in half-metals and strong ferromagnets \cite{longrange}.

\section{Conclusions}
In conclusion, we have proposed a spin-valve made of two layers of
uniform ferromagnetic layers with unequal thicknesses attached to a
superconducting lead from one side and a normal metal layer from
other side. Our results demonstrate an anomalous positive Meissner
effect which can be experimentally probed in the connected normal
metal layer. We have shown that the anomalous Meissner effect
appears due to the dominance of proximity odd-frequency triplet
superconducting correlations. The theoretical proposed structure
here may open new and feasible venues in experiment to study and
investigate the proximity triplet superconducting correlations.

\acknowledgments

M.A. would like to thank G. Sewell for valuable discussions in
numerical parts of this work. K.H. is supported in part by the
ILIR program and
by a grant of supercomputer resources provided by the DOD HPCMP.
J.L. was supported by the COST Action MP-1201 "Novel Functionalities
through Optimized Confinement of Condensate and Fields".

\end{document}